\documentclass{article}
\usepackage{graphicx} % Required for inserting images
\usepackage[letterpaper, portrait, margin=1.5in]{geometry}

\usepackage[ pdftex, plainpages = false, pdfpagelabels,
                 pdfpagemode = UseOutlines,
                 bookmarks,
                 bookmarksopen = true,
                 bookmarksnumbered = true,
                 breaklinks = true,
                 pagebackref=false,
                 colorlinks = true,
                 linkcolor = blue,
                 urlcolor  = blue,
                 citecolor = BrickRed,
                 anchorcolor = green,
                 hyperindex = true,
                 hyperfigures,
                 citecolor = blue
                 ]{hyperref}
                 
\title{Enaction for QBists}
\author{Amanda Gefter}
\date{October 2024}

\begin{document}

\maketitle
\begin{abstract}
    This paper began as a set of notes introducing quantum physicists of the QBist persuasion to enactive theory. Unlike mainstream cognitive science, which views cognition as computations on internal representations of the external world (and thus the mind as \textit{in} the head), the enactive approach sees cognition as adaptive, embodied action. Enaction can ground concepts of experience, agency, knowledge, and normativity—which play key roles in QBist quantum mechanics—in terms consistent with QBism's participatory approach. That's because QBism and enaction both reject an absolute, pregiven subject-object split: for QBism, quantum measurement is the enactment of the subject-object divide; for enaction, cognition is the enactment of that divide. Indeed, each appears to be half of the same story. QBism is a theory of how the world is created and recreated through interactions with an agent, while enaction is a theory of how agents are created and recreated through interactions with the world. Through conversations with QBists and enactivists, these notes evolved into a larger project aimed at unifying QBist physics with enactive cognitive science. Taken together, they offer the possibility of a unified metaphysics—one that brings subject and object, mind and world, back together again.
\end{abstract}
\section{Introduction}
The physicist John Wheeler wrote \cite{1}: “The quantum principle has demolished the view we once had that the universe sits safely “out there,” that we can observe what goes on in it from behind a one-foot-thick slab of plate glass without ourselves being involved in what goes on. We have learned that to observe even so minuscule an object as an electron we have to shatter that slab of glass […] We have to cross out that old word “observer” and replace it by the new word “participator.” In some strange sense the quantum principle tells us that we are dealing with a participatory universe.” Today, the interpretation of quantum mechanics known as QBism \cite{2, 3, 4}—which “originated in one form or other through Wheeler”—embraces the notion of a participatory universe, suggesting that the formalism of quantum mechanics is a “user’s manual” to “help guide [an agent’s] little part and participation in the world’s ongoing creation”\cite{5}.

Shattering the glass—replacing passive observers with active participators— means rejecting the absoluteness of the traditional subject-object divide. But that divide is a key ingredient in the mainstream view of the mind that dominates cognitive science. “Cognitivism,” as the prevailing school is known, conceives of the mind as insulated by the “plate glass” of the skull, passively receiving sensory information from the outside world and constructing neural representations inside the brain. Cognition is construed as computation on those representations. Consciousness itself supposedly emerges from, or perhaps is identical with, those neural representations—though the hard problem of how that happens or to whom such representations appear is left to magic. In brief, cognitivism says that the mind is in the head, with a clear line drawn between the subject experiencing a virtual world on the inside and the unknowable “real” world on the outside, a division inaugurated by Descartes (for the purpose, it’s worth noting, of creating classical physics). The entire process of perception and cognition, in this view, is passive.\footnote[1]{Even currently popular models such as “active inference” or “predictive processing,” which emphasize top-down rather than bottom-up perception, draw a hard line between the representational brain and the external world such that the process of cognition itself remains fundamentally passive. Predictions based on internal models of the world are compared with incoming sensory data; there’s a passive comparison \textit{against} the world rather than active engagement \textit{with }the world.  }

The real culprit here is the concept of a representation; it requires a neat separation between the subject doing the representing and the object being represented, while Wheeler’s “quantum principle” suggests that the two cannot be so neatly divided. As Bohr often put it, “The finite magnitude of the quantum of action prevents altogether a sharp distinction being made between a phenomenon and the agency by which it is observed.” “This was a possibility of which Descartes could not have thought,” Heisenberg wrote \cite[p.81]{6}, “but it makes the sharp separation between the world and the I impossible.”

Cognitivism is not consistent with a participatory universe, and yet it’s through cognitivism that we typically understand notions like “experience,” “agency,” “knowledge,” and “normativity,” which QBism explicitly brings into the quantum story. If QBism isn’t careful to distance those terms from their cognitivist meanings, it can end up embroiled in contradictions. People learning about QBism are apt to be confused if they read QBist notions back into traditional cognitivist terms. What’s needed is an alternative framework that QBism can draw on in defining its terms. Clearly, physicists working in QBism already sense this, which is why they pull inspiration from philosophical schools such as early American pragmatism and phenomenology, where the subject-object divide isn’t pregiven. But the insights of such traditions have now been drawn together into a full-fledged research program in cognitive science, which can flesh out our understanding of things like experience, agents, knowledge, normativity, and even language in truly participatory terms.
\section{The Enactive Approach to Cognition}
In the 1991 book \textit{The Embodied Mind}, Francisco Varela, Evan Thompson, and Eleanor Rosch outlined the three central assumptions of the cognitivist point of view \cite{7}: “The first is that we inhabit a world with particular properties…The second is that we pick up or recover these properties by internally representing them. The third is that there is a separate subjective “we” who does these things.”  They decidedly rejected these assumptions and offered a radical alternative. “We propose as a name the term \textit{enactive} to emphasize the growing conviction that cognition is not the representation of a pregiven world by a pregiven mind but is rather the enactment of a world and a mind on the basis of a history of the variety of actions that a being in the world performs.”  Enaction is the view that “cognition is embodied action.” 

Whereas cognitivism draws from Descartes and analytic philosophy and from the research programs of first-order cybernetics and computer science, enaction draws from the philosophies of Merleau-Ponty and the American pragmatists and from the research programs of second-order cybernetics and dynamical systems theory. Cognitivism deals in propositional knowledge (knowing \textit{that}); enaction deals in embodied sensorimotor knowledge (knowing \textit{how}). Cognitivism starts with an observer standing apart from the world and asks how she can create a model of that world inside her; enaction starts with an agent fully immersed in the world and asks how she can act adaptively in order to keep on living.

Today, enaction (or “enactivism,” as it is sometimes called) has become a thriving alternative to mainstream cognitivism, along with its fellow “E”s: \textit{embodied}, \textit{extended}, \textit{embedded} and \textit{ecological} approaches to mind. All of these push the bounds of the subject-object divide away from the skull, incorporating the body, material tools, the environment, and social practices into cognitive processes—transforming them into proper constituents of what we call the “mind.” But the enactive approach is unique among the E’s in that not only does it allow the subject-object divide to shift according to context, but it reconceptualizes cognition as the \textit{enaction of the subject-object divide itself}.

Two key ideas are necessary for understanding how this happens: living as autopoiesis and cognition as sense-making. Humberto Maturana and Francisco Varela coined the term “autopoiesis” to mean self-creating: they saw this as the essence of biological life \cite{8}. A cell is the most elemental autopoietic unit: it is a network of metabolic processes that create the components of the network itself, including its boundary—for example, the cell membrane—which the cell uses to separate itself from the world.\footnote[2]{While the membrane is the cell’s most visually obvious boundary, there are other equally important boundaries that the cell creates to distinguish itself as an individual, including its cytoskeleton and, as Ezequiel Di Paolo points out, non-topological boundaries, such as CRISPR-Cas reactions, which act as a kind of self-individuating cellular immune system.} By creating its own boundary, the cell deems itself a “thing” in its own right. It grants itself autonomy \cite{9}. It enacts its own subject-object divide.

Ezequiel Di Paolo has emphasized \cite{10, 53} that there’s an inherent tension of autopoietic existence. Autopoiesis wants two things: to make itself and to differentiate itself. But these directly oppose each other. Di Paolo explains it this way. Self-production requires matter and energy, which the system extracts from the environment. To maximize self-production, the system would want to be fully open to the environment so as to take in everything it needs. To maximize self-distinction, the system would want to be fully closed to the environment, so as to protect its identity as a separate unit. It’s the need to navigate this dialectical tension, Di Paolo says, that gives rise to the notion of agency. 

The compromise an autopoietic system strikes between self-production and self-distinction is to regulate its interaction with the environment—sometimes remaining open, sometimes closing itself off—depending on internal needs and external conditions. The cell has a semi-permeable membrane—permeable enough to let in nutrients but solid enough that the cell remains a cell—and now it must decide which to do when. In becoming a decision-maker, the cell becomes a rudimentary agent. 

The process of decision-making entails \textit{sense-making}: the agent must differentiate helpful from harmful, good from bad, not in absolute terms but relative to its current autopoietic needs. The world, as Evan Thompson and Mog Stapleton put it, becomes “a place of salience, meaning, and value” just by virtue of the agent’s being alive \cite{11}. Thus to exist, in enactive terms, is to interact with the world normatively. “An autonomous system produces and sustains its own identity in precarious conditions and thereby establishes a perspective from which interactions with the world acquire a normative status” \cite{11}. “Exchanges with the world are thus inherently significant for the agent, and this is the definitional property of a cognitive system: the creation and appreciation of meaning or sense-making, in short” \cite[p.39]{12}. This direct line from living to sense-making is what enactivists refer to as life-mind continuity.

Living systems can adapt to their circumstances in one of two ways. They can regulate their own internal states through metabolic adjustments, or they can modulate their relation to the rest of the world through action. “Even a bacterium can do more than to just adaptively rearrange its internal metabolic pathways,” the enactivists write \cite[p.9]{9}. “It can actively improve its environmental conditions by seeking out areas with greater concentrations of nutrients both by random search and gradient following.” Sense-making thus leads to sensorimotor engagement. Receptors on a single-celled organism detect chemical signals, which trigger the movements of cilia and flagella, which move the cell through the environment, which change the incoming sensory signals, forming a sensorimotor loop. 

The circular organization of the sensorimotor loop—with sensory signals shaping action and action shaping sensory signals—is key to the enactive view of cognition and perception. By modulating those loops as they go around, an organism is at once behaving more or less intelligently and perceiving more or less successfully depending on how changes in those loops support or hinder autopoiesis. A hungry bacterium successfully navigating to a region with a higher sugar density can be said to be behaving intelligently and accurately perceiving the sugar gradient in its environment without ever having to form an internal model of its world. The same could be said for a human being, whose central nervous system coordinates its thirty trillion or so cells such that the body as a whole can enact its own sensorimotor loops, which both individuate and sustain the body as a single, autopoietic unit.\footnote[3]{There’s some debate about whether the term “autopoiesis” really applies to a multicellular organism like a human being, where there’s not as clear an analogue of the cell membrane, though there are other self-individuating boundaries (see footnote 2). Some enactivists prefer to describe multicellular organisms as “autonomous” rather than “autopoietic.” I’m choosing to stick with “autopoietic” here to emphasize the core enactive ideas and the ways in which they bootstrap complex forms of mindedness from living itself.} In the representational view, information from the senses is corralled into an inner picture of the world to be apprehended by the brain; in the enactive view, the brain participates in the enaction and modulation of sensorimotor loops that allow the body to interact with the environment in such a way as to remain alive. 

There’s a wealth of experimental evidence showing that passive reception of sensory signals does not a perception make.\footnote[4]{See, for instance, Richard Held’s famous “kitten carousel” experiment or Paul Bach y Rita’s sensory substitution experiment (passive vs active cases), to give two classic examples.} To perceive, we must act. But it’s not as if we act and receive perceptions as a result; perceptions \textit{just are} adaptive actions. Psychologist Kevin O’Regan describes enactive visual perception this way \cite[p.29]{13}: “There is no final image. Instead, seeing is considered as something that the person \textit{does}. Seeing is the process of being engaged in asking and answering questions about what is before us, if need be by exploring the changes in visual input that occur as we move our eyes around. There is no end product to be contemplated by a homunculus.”

The result of perception is not an image or movie in the head, but a sensorimotor coordination, a skillful, embodied, adaptive adjustment to the world that an organism makes in order to maintain its identity and autonomy in the face of the entropy.\footnote[5]{In the enactive view, agents are self-producing (autopoietic) systems whose existence is at all times precarious and unfinished—they must act on the world to resist entropy, which is always threatening to undo them, creating themselves again and again in the process.} It’s how it draws and re-draws the subject-object divide. The agent’s perception—being active—reshapes the world and the agent, in turn, must reshape himself to adjust to that new world, which reshapes it again, and so on. The “accuracy” of a perception, in this view, lies not in some (impossible to attain) correspondence between an inner representation and an outer reality but in the resulting success or failure of the body’s coupling with the world, with all of its autopoietic ramifications. Perception, in the enactive approach, is normative. 

What’s the brain’s role in all of this? As the enactivist philosopher Shaun Gallagher writes \cite[p.47]{14}, the brain “\textit{responds} to the world rather than \textit{represents} it. Specifically, it responds not by representing, but through a dynamical participation in a large range of messy adjustments and readjustments that involve internal homeostasis, external appropriation and accommodation, and larger sets of normative practices, all of which have their own structural features that enable specific perception-action loops, that in turn shape the structure and functioning of the nervous system.”

For enactivists, then, the brain alone is simply not the relevant unit of analysis when it comes to cognition. As Evan Thompson put it, looking for the mind in the brain is like looking for flight in a wing. Instead, cognition is constituted by the entire brain-body-environment system and the action-perception loops running through them. “The mind” is a deeply relational notion, existing neither “inside” the subject or “outside” in the world, but in the sensorimotor couplings between them. It is, as Di Paolo says \cite{15}, “an entity made of a relation.”

The brain, of course, is an essential piece of the puzzle, because it wires in past sensorimotor correlations to produce habits of action that can be triggered or modulated by sensory information from the world. (The very first neurons, which evolved in jellyfish-like creatures, linked sensors to muscles. The brain, from its evolutionary origins, was always for wiring up sensorimotor behavior.) Brain structure encodes the “history of the variety of actions that a being in the world performs” (though so do, in many cases, the body and the world—through muscle memory, say, or paths laid down in walking); through synaptic linkages, the brain allows the body to “predict” (through motor preparations, cashed out in behavior) not the structure of a bare world in itself, but the potential results of interactions between the agent and its environment. In that way, the brain allows the body to behave intelligently. But intelligence is not \textit{in} the brain. 

“The brain is an interactive, mediating organ,” says enactivist psychiatrist Thomas Fuchs \cite{16}. “But in the brain itself there is no experience, no consciousness, no thoughts—all this exists only in the interplay of organism and environment.” Likewise enactivists Thompson and Stapleton write \cite{11}, “What goes on strictly inside the head never as such counts as a cognitive process.”

For the enactivist, there can be no such thing as a Boltzmann brain or a brain in a vat.\footnote[6]{That is, a lone brain in a vat, or a disembodied brain that fluctuates into existence after some Poincaré time, would not have experiences, precisely because experiences are not in brains. A disembodied brain is like a disembodied car engine—the engine can’t drive. Experiences run through the dynamics of brain-body-world loops. You need the whole system.} Experience is not, as the philosopher Marjorie Grene put it, a “secret inner something” \cite[p.26]{17}. “Consciousness isn’t something that happens inside us,” writes philosopher Alva Noë \cite[p.24]{18}, “it is something we do.”

Here, then, in brief, is the enactive story: To live is to be autopoietic; to remain autopoietic is to exert agency; to exert agency is to practice sense-making; to practice sense-making is to experience the world in normative terms; to experience the world in normative terms is to be a cognitive, affective, perceiving, sentient creature. It’s not to use a mind to represent a world. It’s to enact a mind and a world through action. 
\section{The Body}
Embodiment is a central feature of the enactive approach. Cognition is the adaptive sensorimotor behavior of the entire living body interacting with its environment in order to maintain its own existence. This is a point that’s easily misunderstood. Take, for instance, the following consequence of enaction: a disembodied artificial intelligence cannot be cognitive; it can’t perceive or have experiences, think or understand, precisely because it doesn’t have a body. “Well, no problem,” a misinformed AI proponent might shrug, “we’ll just stick our AI in a robot and \textit{voila}, it’s embodied.” But that entirely misses the point. It embraces an understanding of the body that goes back to Descartes, who first likened the body to a machine. When it comes to the mind-body problem, it’s not just that we’ve understood “mind” wrong, we’ve misunderstood “body,” too. A robot doesn’t have a body. A robot is an object. A body, on the other hand, \textit{straddles the subject-object divide}.

This is hinted at already in Merleau-Ponty’s notion of the flesh—“Our body is a being of two leaves,” he wrote \cite[p.137]{19}, “its double belongingness to the order of the “object” and to the order of the “subject” reveals to us quite unexpected relations between the two”—but it comes to fruition in autopoiesis. An autopoietic body is not a material object akin to a rock or a robot; it’s an activity—specifically, the activity that makes the material that participates in the activity. Whereas objects are finished entities, bodies are perpetually restless. “Bodies are unfinished,” the enactivists say \cite[p.7]{20}, “always becoming.”

No part of the body is exempt from the frenzy, as it churns itself into being, the human body turning over 330 billion cells every day \cite{21}. Hans Jonas, the German philosopher of biology, described this frenzy beautifully in a footnote\footnote[7]{All of Jonas’s most brilliant moves occur in footnotes.} to his essay “Is God a Mathematician? The Meaning of Metabolism” \cite[p.76]{22}: \begin{quote}
    The exchange of matter with the environment is not a peripheral activity engaged in by a persistent core, it is the total mode of continuity (self-continuation) of the subject of life itself. The metaphor of “inflow and outflow” does not render the radical nature of the fact. In an engine we have inflow of fuel and outflow of waste products, but the machine parts themselves that give passage to this flow do not participate in it: their substance is not involved in the transformations which the fuel undergoes in its passage through them; their physical identity is clearly a matter apart […] Thus the machine persists as a self-identical inert system over against the changing identity of the matter with which it is “fed”; and, we may add, it exists as just the same when there is no feeding at all: it is then the same machine at a standstill. On the other hand, when we call a living body a “metabolizing system,” we must include in the term that the system itself is wholly and continuously a result of its metabolizing activity, and further that none of the “result” ceases to be an object of metabolism while it is also an agent of it. For this reason alone, it is inappropriate to liken the organism to a machine […] Metabolism is the constant becoming of the machine itself—and this becoming itself is a performance of the machine: but for such performance there is no analogue in the world of machines.”\footnote[8]{This view of the body, of course, speaks to the impossibility, within an enactive understanding, of “strong AI” or “general intelligence,” for without autopoiesis and the norms it brings, a machine is not capable of sense-making. Some nuance, however, leaves the door to AI open just a crack. Enactivists speak of “bodies” as existing in a hierarchy of levels: organic, sensorimotor, linguistic, and so on, each one autonomous, thanks to its own processes of self-creation and self-distinction, and yet each rooted in and constrained by the layer of autonomy beneath. Could roboticists create an artificial sensorimotor body that enacts itself as an autonomous agent despite lacking any metabolic autonomy beneath? This remains an open question.}
\end{quote}

The enactivists echo Jonas’s understanding of the body. A body is not something we \textit{have}, they say, which we then use to act in the world; rather, we act in the world and by doing so \textit{achieve} the body we use to take those actions. And because those actions are, by definition, goal-oriented—the body’s goal being to make and remake itself—the body takes a stance on the world. The body \textit{cares}. In making itself an object, the body is also a subject. 

Through autopoiesis, we enact a subject-object distinction. We individuate ourselves \textit{from} the world by the very same move that couples us \textit{to} the world. It is an activity whose result is the body—the mind, the self—but it is never finished, for the moment it stops drawing that line, it disappears.
\section{Knowing \textit{That} vs. Knowing \textit{How}}
According to Gilbert Ryle, knowledge comes in two species: \textit{knowing that} and \textit{knowing how} \cite{23}. Traditional views of philosophy (going back to Plato) and of cognitive science take knowing \textit{that something is the case} to be superior to and more fundamental than knowing \textit{how to do it}. However impressive one’s practical skills, they say, one doesn’t truly understand something until one can describe it intellectually, in terms of sentences, principles, and rules. 

Knowing \textit{that} is a form of representation. One must stand back from an object and represent it as a proposition (“the crow is black”; “it is raining”) whose truth value is determined by its correspondence with the world (the crow is actually black; it is actually raining). Of course, when all knowledge is propositional, there’s no way to know what’s actually the case—one can only know other propositions. This is the so-called “symbol-grounding problem,” which plagued early attempts at artificial intelligence, as AI researchers sought to endow computers with general intelligence by giving them some vast set of propositions. It didn’t work. Propositions were of no use when it came to context and common sense—the first clues that perhaps knowledge isn’t entirely propositional, if it’s propositional at all.

Quick: which finger do you use to type the letter “u”? Answer with your hands tied behind your back. Surely your hands are irrelevant, if your brain need only call up a proposition. And what about knowing how to ride a bike? Is that fully exhausted by some set of statements? Sure, you could try to describe to a child how to ride a bike if they didn’t already know, but surely the ability to describe it is different than the ability to do it, which is itself a kind of muscle memory, an embodied skill. Just ask the kid straddling a wobbly Schwinn, trying to balance on a sentence. 

If knowing \textit{that} is representational, knowing \textit{how} is participatory. It’s a bodily, sensorimotor form of knowledge, but that doesn’t mean it sits inside the body, like a gift inside a box. Knowledge how is not a noun. It’s the body’s readiness to spring into useful action when confronted with a particular context. You may not know how you tie your shoes—not propositionally, anyway—but put a shoelace in your hand and you can do it. Likewise, you probably don’t know the 16th word of the slow song you once danced to at your high school prom, but when it comes on the car radio after all these years, the tune carries you there, prompting one word from your lips, whose movements in turn prompt the next, until you’re belting that 16th word as the light turns green. 

In your reliance on the music to elicit certain bodily responses—certain knowledge \textit{how}—you’re not so different from a blue fin tuna. The philosopher Andy Clark tells the story of how scientists studying blue fin tuna were struck by the fact that, based on the fish’s body alone, it shouldn’t be capable of swimming as fast as it does. In fact, the fish was seven times stronger a swimmer than its muscles would seem to allow. The solution to the mystery was that the tuna flicks its tail to create vortices in the water, then uses those vortices to propel itself forward. “The real swimming machine,” Clark writes \cite[p.143]{24}, “is thus the fish in its proper context: the fish plus the surrounding structures and vortices that it actively creates and then maximally exploits.”

In the enactive perspective, we think the way the tuna swims. The thinking machine is not the brain, not even the brain plus the body, but the organism coupled to the environment in such a way that the organism can give itself a boost into having an even better grip on the world.

There’s a famous thought experiment in philosophy \cite{25} about a neuroscientist named Mary who is locked, from birth, in a monochromatic room where all she ever sees are shades of gray.  Mary (the story goes) knows everything there is to know about the physics of light, the workings of the eye, and the visual system in the brain. She knows everything there is to know, that is, about seeing the color red—except what it’s like to actually see it. One day, she leaves her locked room, stumbles out into the world, and sees a red rose. So, the philosophers ask: does this experience give her new knowledge? If so, then seeing red must be something over and above the facts of physics, a \textit{quale}, the \textit{what-it’s-likeness} of it all, the private, ineffable stuff of subjectivity. That’s one answer. Another is that knowledge isn’t propositional. Knowing redness is not a matter of knowing \textit{that}. It’s an embodied, enacted skill, a way of coupling ourselves to the photon environment, like a tuna to the water, with the movements of our bodies and eyes, with our histories, with our ability to distinguish surfaces, with our linguistic practices, all of which we bring to the act of seeing.

As embodied creatures, we’re constantly trying to position our bodies to resonate with the aspects of the environment we want and ignore those we don’t—we’re working to keep our maximal grip on the world. We put on coats when we’re cold, sweat when we’re hot; when we read, we adjust our arms to bring the text into the right focal distance. We tilt our heads to decipher a menu like plants turning our leaves toward the light. Our sensorimotor engagement with the world is a never-ending series of adjustments and readjustments as we work to maintain that maximal grip, and the power of the brain is not to magically conjure up virtual internal worlds to be seen by who-knows-what, who-knows-how, but to wire up a vast network of correlations so that with every little movement we make, an entire history of adjustments comes trailing along, which we can later—in imagination, fantasy, daydream—subtly replay on our bodies, like plants going through the motions of basking in the sun. 

In the enactive approach, all knowledge is knowledge \textit{how}—whether you’re playing tennis, tap dancing, visualizing a scene, solving a math problem, pondering your existential angst, experiencing the redness of red—it’s knowing \textit{how} all the way up.

Consider logic: so seemingly abstract and removed from the contingencies of bodily life. William James once wrote \cite[p.245-6]{26}, “We ought to say a feeling of \textit{and}, a feeling of \textit{if}, a feeling of \textit{but} […] quite as readily as we say a feeling of \textit{blue} or a feeling of \textit{cold}.”  Of course, he was right. We know what these words mean by their facilitations and modulations of our embodied know-how. \textit{And} is a welcome opening of the arms; \textit{if} is a hesitation to exhale, a searching round of the room; \textit{but} is a palm in the air, a toe stubbed on unnoticed stone, the sudden halting of forward motion. The basic operations of formal logic—\textit{and, if, not}—derive from these lived experiences, even if they seem sedimented out. In the spirit of Charles Sanders Peirce, we might say that logical reasoning is rooted in habit—drawn in sensorimotor loops and bodily dispositions, rooted in the norms of sense-making that keep us coupled to the world and breathing. But habits only serve us until they don’t; they are reshaped each time we enact them, as they confront the incessant novelty of the world. As the philosopher Mark Johnson writes \cite[p.105]{27}, “Nietzsche, James, Dewey, and a host of subsequent thinkers have shown us that life is change and existence is an ongoing process. There is no eternal logic, no absolute form that could save us from grappling with change every moment of our lives.”

The same goes for mathematics \cite{28}. “If our basic modes of understanding were to change, due primarily to changes in our bodies, our brains, or our world, then our mathematics would change also,” Johnson writes \cite[p.107]{27}. “From the perspective of embodied cognition, mathematics remains as beautiful and amazing and stable as it always has been, but it also remains dependent on our ability both to sustain it and to extend it creatively as part of a process of inquiry.”  To paraphrase Noë, mathematics isn’t something inside us, nor is it something out in the world: mathematics is something we \textit{do}. The result of that doing is never a proposition, never a shard of some eternal truth, but simply adaptive know-how we can add to our bodily repertoires, to put to use in the ceaseless flow of our living. 

The most abstract sorts of thinking, then, go on not in our heads, but in our embodied interactions with the world—with the page, with the chalkboard, with our muscular tensions and the motions of our eyes. Even the act of stating a proposition is just another activity, a knowing \textit{how}. “To know is to act,” Di Paolo writes \cite{29}. “Activity does not simply \textit{serve} knowledge; knowing \textit{is} itself an active process. Enaction is a thoroughly \textit{performative} approach to life and mind.”  

The elemental unit of knowledge, then, is not a proposition but a sensorimotor coordination, a bodily resonance with the world that spans the subject-object divide. When we get a successful grip on the world, it implies a kind of knowing, as if we’ve hit on a fact, but it’s a fact that cannot be abstracted away, and is only meaningful in its enaction. To capture the performative aspect of this elemental unit of knowing, Di Paolo coined the term “f/act.” “Each act is thus a fact in the world, a f/act,” he writes \cite[pp. 177-180]{29}. Being enacted rather than propositional, a f/act straddles the subject-object divide. “It is both agent- and world-dependent,” Di Paolo says. “A f/act is an onto/epistemic concept [...] F/acts are open and they can be taken up, resignified by other f/acts [...] Situations are always active, i.e. not just settled facts, but f/acts in various stages of becoming [...] A situation embodies its past. As situations change, so can the sedimented structures with their remaining potentialities change. A past that does not change is not the real past.” 

Building on an epistemology of knowing \textit{how}, enactivists are now adding to Ryle's duo a third form of knowledge, one in which the enaction of the subject-object divide becomes the most ambiguous and therefore, perhaps, the most consequential. The enactivist Hanne De Jaegher calls it an engaged (or engaging) epistemology; we might call it knowing \textit{with}. And it requires a radical shift in perspective. 
\section{First, Third…Second?}
Cognitive scientists and philosophers of mind have long struggled with the irreconcilability of third-person and first-person perspectives. Starting with a third-person world, there’s no clear way to account for the existence of individual first-person perspectives; this is known as the hard problem. But starting in first person, with the mind—conceived in cognitivist terms, as a kind of internal virtual reality run on the hardware of an individual brain—there’s no way to undeniably assert the presence of an external, third-person world. You might call this the solipsism problem.

Quantum theory has struggled with this same dilemma in its own terms. Start with unitary evolution of a universal wavefunction (third person) and there’s no clear way to account for the existence of unique measurement outcomes (first person), but start with an agent’s experience of measurement outcomes and it’s easy to get stuck there, with no obvious link back out to a shared world. 

QBism rejects any third-person description of the world. As Fuchs says \cite{5}, “QBism don’t do third person!”  The grammar is meant to imply a first-person view. But there are already hints within QBism that first person isn’t a valid option, either. True first-person statements are self-referential. They are statements \textit{about oneself}. But such statements are explicitly disallowed in the QBist picture, where it is meaningless to assign a quantum state to oneself.

In philosophy, the “I”—Descartes’ \textit{cogito}—is assumed to have some privileged, private, interior access to itself; it knows itself in a way that is fundamentally different from the way it knows the world or other people. Cognitivists call this introspection. If introspection is possible, then to say “I know that the apple is red” is equivalent to saying, “I know that I know that the apple is red.” For the cognitivist, adding “I know” adds zero content, because our thoughts, beliefs, perceptions and other internal states are, in the first-person perspective, simply given to us in an immediate and transparent way. We simply \textit{have} thoughts, beliefs and perceptions—we don’t have to \textit{do} anything to learn about them. We don’t have to measure them. To put it another way, a true first-person statement does not enact a subject-object divide. All collapses onto the side of the subject and its self-knowing. It’s the mirror-image of a third person statement, which is all object, no subject. The “I” is as much a hidden variable as the “it.”

Bohr himself sensed that there was something problematic about the “I” in quantum mechanics. His favorite book was an unfinished novel called \textit{Adventures of a Danish Student} by Poul Martin Møller; Bohr urged everyone who came within his circle to read it, a rite of passage into Copenhagen quantum mechanics \cite[p.58]{30}. For those who couldn’t manage the language, he would translate the most important part: 

\begin{quote}
    My endless inquiries made it impossible for me to achieve anything. Moreover, I get to think of my own thoughts of the situation in which I find myself. I even think that I think of it, and divide myself into an infinite retrogressive sequence of “I’s” who consider each other. I do not know at which “I” to stop as the actual, and in the moment I stop at one, there is indeed again an “I” which stops at it. I become confused and feel a dizziness, as if I were looking down into a bottomless abyss, and my ponderings result finally in a terrible headache. 
\end{quote}

What lesson about quantum mechanics did Bohr intuit in those lines? Perhaps it was that there’s no way to get at a bare “I”—one always needs a subject-object divide, even when observing oneself, and yet, as Bohr liked to emphasize, the subject-object divide is, in quantum mechanics, fundamentally ambiguous. We need it, but it’s not a given. The old Cartesian \textit{cogito} couldn’t be postulated in the new physics.

For enaction, contra Descartes, the “I” is not the starting point for life or for mind, let alone for philosophy or for science. Enaction is a theory of the ongoing creation of the self through the sensorimotor loops that sustain autopoiesis. But consider an infant. As babies, we can’t feed ourselves, regulate our body temperature, direct our own attention, move our eyes in synchrony, lift our heads, burp. We are not yet fully autonomous sensorimotor agents. Our caregivers have to act in coordination with us to achieve these necessary feats—which is to say, they form part of the sensorimotor loops that enable our continued existence. And yet those sensorimotor loops literally constitute our selves. Our \textit{minds}. The Finnish philosopher Pentti Määttänen writes \cite[p.14]{31}, “Mind is not a property of the brain or even the body. Mind is a property of organism environment interaction as characterized with the loop of perception and action. Any attempt to separate one element from this loop has the consequence that mentality is lost away. Drop the loop, lose the mental.”  If others form part of that loop, they form part of our minds.

Lev Vygotsky \cite[p.56]{32} famously gave the example of an infant grabbing for a toy that is just out of reach. The caregiver, seeing the baby’s intention, grabs the toy and hands it to him, completing the sensorimotor act. The failed reach has given birth to pointing—the baby learns that the gesture directs the caregiver’s attention in a very particular way. With gesture—be it pointing or vocalizing—the child draws the caregiver into the very sensorimotor loops that define the child as a self. Eventually the child can turn those gestures on himself and use them to regulate his own attention and coordinate his own behavior—say, by talking to himself, out loud at first and eventually silently, in a process we call “thinking.” 

Ryle wrote \cite[p.27]{33}: \begin{quote}
    Keeping our thoughts to ourselves is a sophisticated accomplishment. It was not until the Middle Ages that people learned to read without reading aloud. Similarly, a boy has to learn to read aloud before he learns to read under his breath, and to prattle aloud before he prattles to himself. Yet many theorists have supposed that the silence in which most of us have learned to think is a defining property of thought. Plato said that in thinking the soul is talking to itself. But silence, though often convenient, is inessential, as is the restriction of the audience to one recipient. The combination of the two assumptions that theorizing is the primary activity of minds and that theorizing is intrinsically a private, silent or internal operation remains one of the main supports of the dogma of the ghost in the machine. People tend to identify their minds with the ‘place’ where they conduct their secret thoughts. They even come to suppose that there is a special mystery about how we publish our thoughts instead of realizing that we employ a special artifice to keep them to ourselves. 
\end{quote}

“The so-called dialogue with oneself is possible only because of the basic fact of men’s speaking with each other,” wrote Martin Buber \cite[p.112]{34}.   When we do think on our own, it’s not a metaphysically private process, just a set of sensorimotor loops that remain close to home. 

The point is, we start off in life as open sensorimotor loops and others must step in a close them for us, completing the circuit of our self. Only from others do we learn to complete our own circuits—though they always, without exception, run through not only our brains and our bodies but our environments, which are often made up of other people, who participate in our selves to enlarge and enliven them. In enaction, the old hallmarks of the Cartesian self—an individual turned inward, profoundly alone, introspectively reflecting on his own thoughts—are forged externally, in interaction with others. We know ourselves the same way we know anyone and anything else: outwardly, by way of the world, which we act on and then react to the consequences of our actions by taking new actions.  Considering the individual, isolated agent as an ontological primitive, then, will never work. As the pragmatist philosopher George Herbert Mead\footnote[9]{While QBism has taken much inspiration from the pragmatist philosophy of William James, Mead, also a pragmatist, may have a lot to offer, particularly on questions of agent-agent interactions.} put it \cite{35}, “The individual is an other before he is a self.” 

John Dewey agreed \cite[p.170]{36}: \begin{quote}
    When the introspectionist thinks he has withdrawn into a wholly private realm of events disparate in kind from other events, made out of mental stuff, he is only turning his attention to his own soliloquy. And soliloquy is the product and reflex of converse with others; social communication not an effect of soliloquy. If we had not talked with others and they with us, we should never talk to and with ourselves. […] Through speech a person dramatically identifies himself with potential acts and deeds; he plays many roles, not in successive stages of life, but in a contemporaneously enacted drama. Thus mind emerges.
\end{quote}

In enaction, the “I” is not fundamental; in QBism, the “it” is not fundamental. And neither should come as a surprise, since to take either first- or third-person perspectives too seriously is precisely to assume the absolute reality of the subject-object split, which both QBism and enaction deny. “QBism would say, it’s not that the world is built up from stuff on “the outside” as the Greeks would have had it. Nor is it built up from stuff on “the inside” as the idealists, like George Berkeley and Eddington, would have it. Rather, the stuff of the world is in the character of what each of us encounters every living moment—stuff that is neither inside nor outside, but prior to the very notion of a cut between the two at all” \cite{37}.  Likewise, in \textit{The Embodied Mind}: “Our intention is to bypass entirely this local geography of inner versus outer” \cite[p.172]{7}; “There is neither an objective nor subjective pole” \cite[p.225]{7}.

If there’s neither an “I” nor an “it,” neither first person nor third—what’s left? Consider “you.” A word stretched between two poles, between the one who says it and the one to whom it is said, it straddles the subject-object divide. As Buber put it \cite[p.54]{38}, “When one says You, the I of the word pair I-You is said, too.”  Whereas third-person (realist) descriptions deal in objects without subjects, and first-person (idealist) descriptions deal in subjects without objects, a second-person ontology lives in what Buber called the “between” \cite[p.204]{39}.\footnote[10]{“On the far side of the subjective, on this side of the objective, on the narrow ridge where I and Thou meet, there is the realm of “between”.”}

Some enactivists, following Varela and his case for “neurophenomenology,” talk about a kind of circularity through which we swing from first person to third person and back again in a kind of strange loop. I suspect this urge comes from a too-classical way of thinking about the physical world, an assumption that there \textit{is} some third-person reality to contend with and accommodate in an enactive story. In QBism, there is no such thing. In letting that go, it becomes clear that the actual enaction of self and world is something that happens in second person.

Second person is a genuinely participatory perspective. Instead of starting with a fixed boundary in the middle and then trying to figure out how one could ever transcend it, you start in the Between, \textit{then enact the boundary through the interaction itself}. Doing so, Buber says, is always a struggle—you’re vulnerable, because, as the enactivists stress, the very boundaries of your self are at stake—but in doing so you create something new. 

“I think the greatest lesson quantum theory holds for us,” writes Fuchs \cite[p.9]{40}, “is that when two pieces of the world come together, they give birth.”  With a fixed subject-object split, there can be no novelty; all knowledge becomes propositional; the “I” and the “other” are doomed to ping-pong those propositions endlessly back and forth in a sterile block universe haunted by a hard problem and a measurement problem, neither of which can ever be solved. In second person, everything is different. But you can only get there by making two moves: by getting the world out of third person (as QBism does) \textit{and} by getting the self out of first person (as enaction does). Only then, in second person, can genuine novelty emerge. 
\section{The Problem of Multiple Agents}
“What happens when several observers are “working” on the same universe?” John Wheeler asked himself in his journal \cite[entry dated 27 January 1974]{41}. For decades he agonized over this: How can a participatory universe also be a shared universe? “How are acts of observer-participancy fitted together?” he wrote \cite[entry dated 18 April 1980]{41}. “That’s the central mystery…Is there any point partway between all and none on this issue? Each of us a private universe? Preposterous! Each of us see the same universe? Also preposterous!”

The point partway between all and none—between third person and first—is second person. In a cognitivist account, an interaction between two observers results in isolated internal representations—two private, secret inner somethings—with nothing truly shared. In an enactive account, interactions happen in the between. 

Hanne De Jaegher and Ezequiel Di Paolo formalized an enactive approach to intersubjectivity in their theory of participatory sense-making \cite{42}.  The idea is that the same dynamic processes of sense-making through which individual agents create and individuate themselves can also take place in social interactions \textit{between} agents. 

Social interactions (as opposed to, say, interactions with non-living objects, or what Buber would describe as “I-it”) involve two agents actively regulating their engagement. (Two people accidentally bumping one another on the street is an I-it interaction; their eye contact, utterances of apology or annoyance, the quick reestablishment of appropriate interpersonal distance, one reaching out to help the other up, etc., comprise a true social interaction, or I-You.) In social interactions, agents coordinate movements and utterances (often eye blinks and, at times, respiratory and heart rates) in turn-taking rhythms of vocal and bodily behavior. These coordinations create, individuate, and sustain the shared interaction, granting it a temporary autonomy analogous to the autopoietic autonomies of the individual agents. With autonomy comes normativity, as situations and behaviors become good or bad relative to how helpful or harmful they are to continuation of the interaction, which is always at risk of breaking down. That vulnerability is the interaction’s most crucial ingredient. As in the autopoietic case, it’s the looming possibility of breakdown that requires the autonomous system to constantly recreate itself and imposes the norms the system must navigate in order to continue on. 

Though the interaction takes on its own autonomy, it cannot negate the autonomies of the individual agents without putting an end to the interaction. It can, and often must, modulate those autonomies, but it can’t deny them altogether. To participate in a waltz, I have to relinquish some autonomy to allow my body to be led; to participate in a conversation, I have to inhibit my urge to speak until it’s my turn. But if I go limp like a rag doll, or if I stop speaking entirely, the dance, or the conversation, can’t continue. My partner can momentarily pick up the slack—moving my body into a new position, asking me a question to prompt me to speak—and perhaps together we can recover the interaction. But if I fail to reengage, there’s no dance; the conversation turns to monologue; the I-you relation devolves to I-it; the interaction is over.

When, however, interactions persist, “new domains of social sense-making can be generated that were not available to each individual on her own” \cite{42}.  This is the crucial difference between the representationalist and enactive views of social interaction: in the representational story, where each agent is passively modeling the other, nothing substantially new can result from the interaction that isn’t already there in the individual actors. In participatory sense-making, by contrast, genuine novelty can arise. Two agents can know things together that neither knows alone. As always, this knowledge is knowledge-\textit{how}; it’s not propositional. “Shared know-how does not amount to the sum of the individuals’ know-hows nor does it strictly “belong” to any of the participants. It involves instead the practice of coordinating sensorimotor schemes together, navigating breakdowns, and it belongs to the system the participants bring forth together” \cite[p.75]{20}.  Cognitive processes—thinking, perceiving, knowing—which are usually taken to be metaphysically private, can, in participatory sense-making, take place between people, in second person.

It's important to note, though, that conflicts can and do arise between different layers of sense-making. Di Paolo offers the example of smoking, a sensorimotor habit that conflicts with the ultimate aims of metabolic sense-making. On the flipside, he says, a person can decide to stick to a diet even when her cells are demanding food. Social interactions, too, can work against the individual agents’ best intentions. Di Paolo and De Jaegher point out that in a so-called “narrow corridor dance,” where two people trying to pass one another get stuck stepping side to side in synchrony, the dynamics of the interaction itself frustrate the agents’ individual desires to escape it. To deal with such conflicts, agents can turn to language \cite{20}. 

The simultaneous demands of individual autonomy and interactive autonomy in the course of participatory sense-making gives rise to what the enactivists call a “primordial tension” \cite[p.140]{20}.  Just as the primordial tension between the cell’s autopoietic need to be both open and closed to its environment gives rise to agency, the primordial tension between a participatory sense-maker’s need to be both open to and closed to a social interaction gives rise, in the enactive view, to language. Through other-directed gestures and utterances (which can, in turn, become self-directed gestures and utterances), agents regulate social interactions. (One participant can, for instance, put an end to the “narrow corridor dance” by saying, “You go first.”)

In the enactive perspective, language doesn’t \textit{represent} things. Language \textit{does} things. Languaging (to use the enactivist’s verbification) is a form of participatory sense-making, and it is always a form of embodied action. As such, we can only get at the meaning of words and gestures by enacting them. It’s not as if the meanings come first; rather, in the act of our languaging, words and gestures come to mean things. Consider again Vygotsky’s story of pointing. At first the child is simply reaching for the toy, but when his reach induces the caregiver to hand it to him, the action becomes a gesture—it becomes \textit{meaningful}. The next time he wants a toy, the child can point, regulating the interaction between himself and the caregiver. If the caregiver doesn’t understand the gesture, the child might cry in frustration; a pure bodily response. But if the caregiver recognizes the child’s intention in the cry, the cry, too, can become meaningful, and can be used communicatively in the future to steer the interaction.

The mainstream representational view of language—where words have set meanings from the start—is nothing but the fossilized remains of real, living language. “A symbol is nothing but the stimulus whose response is given in advance,” Mead said \cite[p.181]{43}; in other words, a sensorimotor habit.  Ultimately, though, those meanings can never completely be fixed, or language would cease to be able to regulate participatory interactions.

“Language does not simply symbolize a situation or object which is already there in advance; it makes possible the existence or the appearance of that situation or object,” Mead wrote \cite{43}. Likewise, Merleau-Ponty \cite[p.408-9]{44}: \begin{quote}
    Speech cannot be considered as mere clothing for thought, nor expression as the translation of a signification, already clear for itself, into an arbitrary system of signs [...] Communication certainly presupposes a system of correspondences, such as those given by the dictionary, but it goes beyond, and it is the sentence that gives each word its sense, it is for having been employed in different contexts that the word gradually takes on a sense that is impossible to fix absolutely. An important speech or a great novel imposes its sense. And as for the speaking subject, the act of expression must allow even the subject himself to transcend what he had previously thought, and he must find in his own words more than he thought he had put there, otherwise we would never see thought, even when isolated, seek out expression with such perseverance. Thus, speech is this paradoxical operation in which—by means of words whose sense is given and by means of already available significations—we attempt to catch up with an intention that in principle goes beyond them and modifies them in the final analysis, itself establishing the sense of the words by which it expresses itself.
\end{quote}

Two adults conversing with well-worn language must still enact new meanings in the course of their interaction if it’s to be a genuine conversation (and not, say, the rehearsed play-acting of small talk). One person speaks; the other responds, “Oh, you mean…?” “Not exactly,” says the first, then reformulates his words, altering, ever so slightly, the original meaning. “Ah, so it’s like…” the other concedes, only in doing so he’s tweaked the meaning again. “I suppose it is,” replies the first. “It’s as if….” By the end of the conversation, a new meaning has arisen, a novel outcome that belongs to the both of them. Unlike the representational view, where communication entails the successful passage of a message from one individual to another, enactive languaging is about the negotiation of new meanings in interaction.

Indeed, if I were to speak words at you, and you were to receive them with no friction, no negotiation whatsoever, I would walk away feeling utterly misunderstood. And yet, after a negotiation, when I \textit{do} feel understood, it’s not because I’ve handed over my thoughts unscathed. It’s because I’ve had to alter them in the process as you’ve recast them in your own words. The feeling of being “understood,” then, is really a kind of basking in the glow of a successful creation, of the \textit{shared} meaning that was forged in the vulnerable space of our between. There Buber emphasizes the literal \textit{responsibility} we have toward one another—a willingness, that is, to respond, to participate, to shape and be shaped by the shared interaction—an interaction that can, in the enactive view, rightly be seen as \textit{thinking together}.

Buber writes:
\begin{quote}
    In a real conversation (that is, not one whose individual parts have been preconcerted, but one which is completely spontaneous, in which each speaks directly to his partner and calls forth his unpredictable reply), a real lesson (that is, neither a routine repetition nor a lesson whose findings the teacher knows before he starts, but one which develops in mutual surprise), a real embrace and not one of mere habit, a real duel and not a mere game—in all these what is essential does not take place in each of the participants or in a neutral world which includes the two and all other things; but it takes place between them in the most precise sense, as it were in a dimension which is accessible only to them both. Something happens to me—that is a fact which can be exactly distributed between the world and the soul, between an “outer” event and an “inner” impression. But if I and another come up against one another […] the sum does not exactly divide, there is a remainder, somewhere, where the souls end and the world has not yet begun, and this remainder is what is essential \cite[p.203-4]{45}. 
\end{quote}

That remainder—the irreducible overlap between subject and object, or perhaps between subject and subject—is reminiscent of the \textit{h} in quantum mechanics, the quantum of action, which also can be said to quantify a subject-object overlap, a remainder, in the course of a measurement interaction. And indeed, the enactive view of language is relevant to quantum mechanics in response to Bohr’s insistence on the use of “unambiguous language” to communicate measurement results. Bohr saw that quantum mechanics raised a problem of intersubjective agreement—\textit{What happens when several observers are “working” on the same universe?}—but believed it could be solved through an appeal to “unambiguous,” “ordinary” language.

“By ordinary language we mean such use of words where a sharp separation between subject and object can be maintained,” Bohr wrote \cite[p.8]{46}. He saw that quantum mechanics itself makes the subject-object divide fundamentally ambiguous, but hoped that language could somehow stand outside the flux, that it could pin down measurement outcomes such that they could be passed around like hidden variables, neatly packaged in propositions, from one observer to another, participation-free. What’s clear in the enactive view is that language doesn’t work that way. Languaging is yet another way we enact a subject-object distinction, another way we participate in the ongoing creation of ourselves and the world. And it’s down to the precariousness of our participatory interactions—down to their deep-rooted ambiguity—that we are able to make meaning at all. 

Thus the enactivists agrees with Buber when he writes \cite{34}, “Language by its nature is a system of possible tensions […] If the tension between what each means […] becomes too great, there arises a misunderstanding that can mount to destruction. But below the critical point the tension need by no means remain inoperative; it can become fruitful, it always becomes fruitful where, out of understanding each other, genuine dialogue unfolds. From this it follows that it is not the unambiguity of a word but its ambiguity that constitutes living language.” 

Ambiguity demands participation and it’s through participation that something new is created in the between, in that little remainder that can’t be divvied up into subject and object. How then does enaction confront the problem of intersubjective agreement? 

“Agreements are not givens, but f/acts themselves,” Di Paolo writes \cite[p.182]{29}. “They must be achieved.”  They are achieved through participatory sense-making, not by compelling two agents to agree on some pre-existing facts in the world, but by allowing two agents to forge new “f/acts” through second-person interactions that are always provisional, always renegotiable, always vulnerable to breakdown, and as such genuine and real. Agreement is never guaranteed, but it can be enacted.

\section{QBism and Enaction}
Why do QBists need enaction? In telling a participatory story of quantum mechanics, QBism must draw on terms that aren’t traditionally part of the physics lexicon. Experience, agency, beliefs, normativity—these are terms that more typically belong to cognitive science. And yet, if we define them using mainstream cognitive science, which draws from the same Cartesian split that participatory quantum mechanics rejects from the start, contradictions are bound to arise. 

(Of course, this is a wider problem in quantum mechanics. If we look, for instance, at non-QBist discussions of extended Wigner’s friend scenarios, it’s clear that first-person, Cartesian ideas about knowledge, perception, language and communication implicitly abound. Words are taken to represent things (as when the friend reports back to Wigner what she saw), and knowledge is packaged in propositions that we carry around in our pockets (“the electron is spin up”) like hidden variables we can pass off to other agents.)) 

Consider, for example, the oft-hurled accusation that QBism amounts to solipsism, a reaction to its claims that quantum states are degrees of belief and that measurement outcomes are experiences for the agent. Are such claims solipsistic? It depends on what one means by “belief,” “experience,” or “agent.” In the cognitivist framework, yes, they are solipsistic: beliefs, in a representational theory of mind, are internal states that can be entirely divorced from action, while experiences are metaphysically private happenings that take place inside the brain and agents are Cartesian cogitos, passively watching the world in virtual reality simulation from within the cozy comforts of the skull. In an enactive framework, things look different. Beliefs are enacted in precarious engagements with the world, experiences are ongoing processes of sense-making through sensorimotor interaction, and agents are embodied, living systems actively coupling to the world in order to survive. They’re not solipsistic in the slightest.

Enaction provides clear answers to QBism’s critics. You think QBism is solipsistic? See enaction: beliefs aren’t representational; experiences aren’t private. You want to know how two agents can agree on the truth value of a proposition? See enaction: knowledge isn’t propositional. You think QBism amounts to idealism? See enaction: mind is embodied and world-involving through and through. It’s not that QBism doesn’t know these things already. QBists themselves make it clear that “QBism does not treat measurement outcomes as propositions” \cite[p.5]{47}  and that “QBism indeed regards agents as embodied; how could a disembodied entity take physical actions and experience consequences?” \cite[p.9]{47}. But enaction provides a rich, well-developed research program and an ever-growing body of literature that QBists can use to back up such claims. 

Enaction may also lead QBists to new insights, pushing the quantum story forward. For instance, people encountering QBism for the first time, with its emphasis on agents and decision-making, are inclined to demand, “But what’s an agent?” At times QBism answers that the agent is a primitive in the theory; by treating quantum theory as an addition to Bayesian decision theory, QBism no more must define an “agent” than classical decision theory did. It’s a fair point, but not exactly satisfying. Here, enaction can swoop in and pick up the story where QBism leaves it—after all, the enactive approach is a theory of agency. And not only is it a theory of what agents \textit{are}, it’s a theory of how agents \textit{come to be}. 

As it happens, though, QBism does have more to say on agency, and again it intersects intriguingly with enaction. When, for instance, contrasting themselves with proponents of Relational Quantum Mechanics (RQM), QBists point out that for RQM, any physical system can be an observer—an electron, a detector, a chair—which is not the case for QBism. What, then, is QBism’s criteria for agency? Fuchs offers this \cite[p.6]{48}: “An agent is an entity that can freely take actions on parts of the world external to itself and for which the consequences of its actions matter to it."

Already we’ve departed from traditional, mechanistic physics, where no one speaks of anything \textit{mattering} to anyone. In QBism, what matters \textit{matters}, because QBism views quantum theory as a decision theory. If one didn’t care about the outcomes of their actions, they simply wouldn’t use a decision theory, and the Born Rule—which in QBism is a normative constraint—wouldn’t hold any power. An agent for whom nothing mattered would be willing to be “Dutch booked,” willing to lose it all as he gambled on the outcomes of his measurements. 

Is there some deeper source of this mattering that goes beyond the desire not to lose one’s shirt playing the quantum ponies? Enaction is a theory of \textit{why things matter}. According to the enactive approach, to exist—or rather, to persist—is to interact with the world normatively. (As the enactivists put it, “Mind is possible because a body is always a decaying body” \cite[p.42]{49}.) It’s the agent’s autopoietic imperative: we’re betting on our lives. And our bets always involve beliefs about the outcomes of multiple, mutually-exclusive actions we could take on the world—do we remain open to the world to aide in self-creation, or do we close ourselves off to self-individuate? Do we engage this sensorimotor habit or that one? The moment we stop placing bets—the moment we decide the outcomes don’t matter—is the moment our lives and our agency come to an end. In enaction, then, it’s normative all the way the down. We have to act to avoid a sure loss—namely, death—in order to create and maintain ourselves as agents. So while QBism starts with agents as primitives, then has them strive for consistency in their beliefs before giving them the green light to act on the world, enaction starts with those conflicting beliefs, which compel actions on the world, which in turn \textit{constitute} the agent. 

Given this view of agents, consider the problem of intersubjectivity in quantum mechanics. Without a deterministic block universe to provide a shared reality from the start, how can participators ever come to agree on the facts of the world? “In QBism,” writes Rüdiger Schack \cite{50}, “intersubjective agreement is not an automatic consequence of the quantum formalism, but a goal that agents might strive for.”  It’s a statement that resonates with the enactive approach: “Agreements are not givens […] They must be achieved” \cite[p.182]{29}. In enaction, they are achieved through participatory sense-making. Can this be useful to the QBist? In participatory sense-making, the interaction between agents has its own autonomy that co-exists along with the autonomies of the individual agents. In other words, the same process by which the individual forges his own identity takes place at the level of the interaction. Could something similar occur between agents in quantum mechanics? 

In QBism, an individual agent must strive to achieve consistency among her own beliefs, which she does by ensuring they conform to the Born Rule. Could, then, \textit{two} agents interacting likewise strive to achieve agreement by making their \textit{combined} beliefs conform to the Born Rule? If so, participatory sense-making would suggest that while the combined beliefs do not have to be identical to either of the individuals’ beliefs, and while they might inspire the individuals to modify their own personal beliefs in the process, they cannot do so in such a way that the individual’s beliefs no longer conform to the Born Rule. (After all, disregarding the Born Rule would mean that the outcomes of their actions no longer matter to them, which would disqualify them as agents.) In other words, the normative constraint of the Born Rule would have to apply at \textit{both} levels simultaneously, otherwise the agreement would no longer be among two agents, and intersubjectivity would lose its “inter.”

Whether this move would get us somewhere is hard to say, but the point is, it’s only in an enactive framework that we can even pose such a suggestion. For if we took agents to be forming metaphysically private internal models of the world, then two agents interacting with one another couldn’t possibly produce a “third agent” between them. Genuine intersubjectivity would be impossible from the start.

Conversations between enactivists and QBists are just beginning \cite{51}, and it’s impossible to say where they may lead.  But according to both research programs, their interaction is bound to produce something novel. If QBism is ultimately a theory of how the world is created and recreated through interactions with an agent, enaction is a theory of how the agent is created and recreated through interactions with the world. One can’t help but feel that each is half of the same participatory story.

\section{Conclusion: Toward a Unified Approach}
“From what deeper principle does the necessity derive of the quantum in the construction of the world?” Wheeler asked \cite[p.97]{52}. “We are first able to play the game when with chalk we have drawn a line across the empty courtyard. It does not much matter where. In quantum mechanics it does not much matter where we draw the line between observing equipment and system observed. Nevertheless, the line must be drawn.” 

QBism and enaction are theories of how Wheeler’s chalk line—the subject-object divide, which was, in classical physics, pregiven and absolute—is enacted and renegotiated interaction by interaction. Both theories reject representationalism, which requires that fixed divide. (“We propose as a name the term enactive to emphasize the growing conviction that cognition is not the representation of a pregiven world by a pregiven mind” \cite{7}. “One of the main conclusions of QBism is that quantum mechanics is not representational” \cite{50}.) For QBism, quantum measurement is the enactment of that boundary; for enaction, cognition is the enactment of that boundary; both are processes that are forever unfinished. For the former, the world is created on the fly; for the latter, the mind is created on the fly. QBism chipped away at the classical object, while enaction chipped away at the classical subject, and between them they opened a new place in the middle—in the drawing of the divide itself—for them to meet.

Today, both theories are still seeking their ontologies, and it’s for that, perhaps, that they need one another most. QBism accounts for the participatory becoming of the world, but often leaves agents as primitives in the theory; enaction accounts for the participatory becoming of agents, but often leaves the world as a primitive. Just as QBism would fall into incoherence if it understood agency, belief, experience, and normativity in representational, cognitivist terms, enaction would fall into incoherence if it understood agents to be acting within a world governed by classical, non-participatory physics. Enaction needs the QBist understanding of the world as crucially as QBism needs the enactive understanding of the agent. (Perhaps, then, these notes should be followed by a companion set: QBism for Enactivists.)

“The classical representationalist picture of thought and how the mind works is so powerful and so deeply rooted in our self-understanding that it is hardly likely to ever be dislodged,” Johnson writes \cite[p.132]{27}. The same, surely, can be said of the classical realist picture of matter and how the universe works. QBism and enaction both stand as challenges to our most deeply entrenched notions of world and self, the very foundations of reality. They face uphill battles, to say the least. But the point is, it’s the \textit{same} battle. Both are fighting against the same 17th century Cartesian split, the same dichotomy of subject and object. Trying to convince others of QBism or the enactive approach in isolation may be tough going precisely because it leaves half the story untold. Together, though, QBism and enaction make for powerful allies, as they can offer, for the first time, a unified post-Cartesian metaphysics, one that finally brings self and world back together again.

\section*{Acknowledgements}
First and foremost, thanks to Christopher Fuchs and the entire QBism group for allowing me to follow them around the world, crash their Zoom meetings, and witness firsthand the incredible creativity, rigor, vulnerability, and insight that can emerge in deeply participatory physics. Special thanks, in particular, to Chris Fuchs, John DeBrota, Marcus Appleby, Jacques Pienaar and Rüdiger Schack for enlightening discussions about the connections between enaction and QBist metaphysics and the meaning of quantum physics in second person. Immense appreciation to enactivists Ezequiel Di Paolo and Hanne De Jaegher for profound inspiration, for engaging so openly and enthusiastically with QBist ideas, and for reading these notes and offering invaluable feedback. Finally, I am deeply grateful to Chris Fuchs and John DeBrota for encouraging me to write up and share these ideas, and to see where the connections between QBist and enactive thinking can lead.

\end{document}